# Ranking library materials

Dirk Lewandowski

Hamburg University of Applied Sciences, Faculty Design, Media and Information, Department Information, Berliner Tor 5, D – 20249 Hamburg, Germany

E-Mail: dirk.lewandowski@haw-hamburg.de

**Abstract**

Purpose: This paper discusses ranking factors suitable for library materials and shows that ranking in general is a complex process and that ranking for library materials requires a variety of techniques.

Design/methodology/approach: The relevant literature is reviewed to provide a systematic overview of suitable ranking factors. The discussion is based on an overview of ranking factors used in Web search engines.

Findings: While there are a wide variety of ranking factors applicable to library materials, today's library systems use only some of them. When designing a ranking component for the library catalogue, an individual weighting of applicable factors is necessary.

Research limitations/applications: While this article discusses different factors, no particular ranking formula is given. However, this article presents the argument that such a formula must always be individual to a certain use case.

Practical implications: The factors presented can be considered when designing a ranking component for a library's search system or when discussing such a project with an ILS vendor.

Originality/value: This paper is original in that it is the first to systematically discuss ranking of library materials based on the main factors used by Web search engines.

Paper type: Conceptual paper

**Introduction**

Most information systems today use some kind of ranking system to bring order to their results lists, which contain large amounts of data. This leads to a situation in which a user is not willing or able to review all the results found. Secondly, users are now used to good relevance ranking, e.g. in Web search engines. These produce at least some relevant results, even for very broad or unspecific queries. When someone is used to such a relevance ranking from general-purpose Web search engines, he or she cannot understand why the search experience at libraries is in general inferior, and the search results in particular are not as good as should be expected from institutions that focus on quality-controlled information, i.e. the best information available.

In this article, I will focus on ranking for library materials. I am certain that the future of the library catalogue is dependent in large part on its ability to produce good results lists, but also the future of the library itself depends on its search systems. I think that this future will be determined by the ability of the catalogue to produce relevant results, even if a user's query is very broad or vague.

This article shows that ranking in general is a complex process. This holds true for search engines as well as for the libraries' OPACs. It demonstrates how search engines operate and how their methods could be applied to the library catalogue.

Before ranking material in the library catalogue, one must first discuss what is really meant by "library material." While today's OPACs mainly index "bundles" (i.e., books, journals), it is crucial to make *all* the material available *in or through* the library searchable within one system. Most users do not understand the distinction between library-controlled resources (catalogue, local digital repositories, course management systems, and institutional Web sites) and remote resources (abstracting and indexing databases, e-journal collections, subject gateways) (Sadeh, 2007), p.310), and, more important, most users have no interest in the subject. All these materials should simply be searchable in the library's information system.

This wider approach to the library search system leads to something similar to a "library search engine" or "academic search engine" than to the traditional catalogue (Lewandowski, 2006; Lewandowski & Mayr, 2006). It is quite interesting that in this regard, commercial operations outside the library sector have far better offerings today. When scientific search engines such as Google Scholar become the standard starting point for academic searches, it becomes clear that search moves away from local offerings to an approach where search comes first, followed by availability. However, the problem with searching systems that are not in the control of the library lies in their quality. and their limited approach regarding what content will get indexed (Lewandowski & Mayr, 2006).

It is important to apply ranking to library materials because of users' expectations concerning information systems in general. We can see that the characteristics of user behaviour in Web search engines (Höchstötter, 2007; Höchstötter & Koch, 2008) now also apply to the library catalogue (see, e.g., (R. Schneider, 2009). Furthermore, user behaviour in the scientific context dramatically changed within the last few years (Rowlands et al., 2008).

The aim of this article is to show what factors could be used to rank the different materials available through the library. The next section provides a review of the main problems with library catalogues today and discusses the reactions from the library community. Then, there is an examination of ranking in Web search engines which shows that the major factors used in this context may be also useful in the library context. The next section analyzes their applicability to the library catalogue. Then, it presents an argument that a good results list should provide a good mix of results rather than "more of the same". The article ends with some concluding remarks and suggestions for further research.

**What exactly is wrong with the library catalogue?**

When one reads the literature and talks to colleagues, it soon becomes clear that something is wrong with the existing library catalogue. However, it is often unclear *what* the problem really is. I think that we can identify four points that make searching the catalogue a disappointing enterprise.

- The catalogue is incomplete. Often, journal articles are missing, or it is unclear to what extent they are covered. In general, the catalogue does not cover all the items available through the library. Therefore, libraries should widen the focus of their catalogues to create something closer to a library search engine than an OPAC (Lewandowski, 2006; Lewandowski & Mayr, 2006).
- Secondly, the catalogues still follow the metaphor of the card catalogue (Breeding, 2006), and searching methods have remained the same.
- Thirdly, user behaviour has changed dramatically in recent years. Search engines have influenced users' demands in that many users now expect to be able to use the techniques they apply in Web searches to other information systems. This results in short, unspecific queries.
- Finally, OPACs should fit both known-item searches and topic-based searches. While in Web search it soon became clear that users use the engines for different query types (Broder, 2002), library catalogues are still focussing on a single query type, namely the informational (topic-based) type. While known-items searches lead to satisfying results when the item is *exactly* known, it is difficult to find a certain item when the title or author is not remembered exactly.

There is no shortage of criticism of libraries' catalogues. The most popular critic may be Karen Schneider (K. G. Schneider, 2006). However, the proposed solutions are not satisfying. The focus is on providing additional features and making the catalogue more "2.0", i.e., adding features for collaborative work and opening the catalogue to other applications not necessarily provided by the library.

These are the general assumptions concerning this new generation of library catalogues (or more precisely, interfaces for the library catalogues).

- Users should participate in creating metadata such as reviews and ratings.
- Generally, the metadata of individual items should be enriched by reviews, tables of contents, and so on.
- Navigation should be improved. This can be achieved by offering drill-down menus on the results pages. The basic ambition here is to combine elements of searching (the initial query) with elements of browsing (reducing the results set by clicking through categories).
- Finally, additional collections owned or offered by the library should also be searched. The approach followed here is usually a federated search (see, e.g., (Joint, 2009).

While all of these features can improve the user's experience and add value to the catalogue, searching is still the core of the application. Unfortunately, improvements to core search techniques in library catalogues have not been significant in recent years. It seems that libraries try to win back users through providing additional features. However, these will be useful only when the search application works well.

Most library OPACs still apply freshness as the only ranking criterion. If a "real" ranking is applied, it usually uses only standard text matching (see, e.g., (Dellit & Boston, 2007). There are some ideas on improving the catalogue through the use of ranking that go beyond pure text-matching. Using popularity ranking factors has been proposed (e.g, (Flimm, 2007)), as well as using circulation statistics, book review data, the number of downloads, and the number of print copies owned by the institutions (Mercun & Zumer, 2008; Sadeh, 2007). However, there is, at least to my knowledge, no systematic overview of suitable factors.

Before going into detail on the ranking factors, some misconceptions about relevance ranking should be dispelled. One such misconception is that a clear sorting criterion is better than relevance ranking. However, one should note that relevance ranking does not reduce the number of results but only puts them in a certain order. Additional searching options could always be provided. So a library catalogue applying ranking must by no means be limited in search options. In contrast, such a catalogue can be a good solution for the inexperienced user and the information professional as well.

The second misconception is that traditional library catalogues do not apply any form of ranking. This is not true, as they order results by publication date. This may be a very simple ranking system, and it assumes that the newest titles are the most suitable, but it still ranks the results. However, while it could sometimes be assumed that current results are the most suitable, this is certainly not always the case. For instance, a simple search for the philosopher Immanuel Kant in the current environment will produce current literature on his works instead of his major works, which are here assumed to be the best suitable results.

To some, relevance ranking may seem a useless enterprise. It is true that "relevance" is hard to define and that it depends on the individual user and the context of the search (Mizzaro, 1997; Saracevic, 2007). However, even a slightly sophisticated relevance ranking can at least produce sufficient results lists. Where known-item searches are concerned, relevance ranking is the only way to display the one desired result in the first position.

The fourth – and by far the worst – misconception regarding relevance ranking is that ranking is easy. It is assumed that one must only apply some standard measures, such as term frequency, to achieve accurate results. This is not true. Good ranking algorithms go far beyond simple text matching.

**Ranking in Web search engines**

Before we discuss the ranking factors applicable to this catalogue, we will look at how Web search engines rank documents. They have the most advanced ranking algorithms, and they also apply specific measures for their context (the Web). Ranking in Web search engines can be used as a prototype for discussing ranking in the library catalogue.

Ranking in Web search engines is based on four groups of factors. There may be a hundred or even more individual factors[i], but all of these will fall into one of the following groups:

- Text matching: This basic form of ranking is measured by how well the text of a certain document matches the query. Among other considerations, term frequency and the position of the search terms in the document are taken into account. (For details, see (Lewandowski, 2005). Search engines use text from the documents themselves and from anchor texts for their analysis.
- Popularity: Search engines measure the popularity of every document they detect. Most often, popularity is measured through linkage, the most popular link-based algorithm being Google's PageRank.
- Freshness: Sometimes, fresh documents are very useful. However, the problem lies in determining when this may be the case. Additionally, date information is hard to extract from Web documents (Lewandowski, 2004).
- Locality: Search engines assume that documents that are "near" the user are more suitable for his or her queries. Therefore, users in different locales may get the same results but in a different order when searching for the same words in search boxes for different countries.

Some of the factors mentioned here are query-dependent, and some are query-independent. (See (Lewandowski, 2005). This is an important distinction, as all query-dependent factors must be calculated "on the fly" (i.e., when the query is sent to the search engine) while query-independent measures can be pre-processed and must be updated only in a pre-determined frequency. Because the response time of an information system is crucial, one must always balance heavy use of query-dependent factors with available processing power.

In particular, the popularity ranking is used to determine a certain quality of the documents in the first positions of the results list. As users in the library context are in many cases not willing or even able to look through the whole results set, quality ranking becomes a crucial factor here, too (Lewandowski, 2008). As will later be shown, in addition to a good ranking, a good mixture of results is also needed.

**Ranking factors for the library catalogue**

In this section, the above-mentioned groups of factors and their applicability to library materials will be discussed. A list of all ranking factors discussed in the text can be found in the appendix.

*Text matching*

Text matching can be applied using such standard measures as term frequency and inverted document frequency. In addition, fields such as title or subject headings can be weighted. These factors are currently used in library catalogues, but many other factors could be applied to improve this basic ranking.

Regarding text, the availability of a small amount of text can be used as a ranking factor. It may be a good idea to prefer documents that allow the user to see the full text or at least a review or an abstract.

The main problem with text matching is that metadata in the catalogue often does not include enough text to achieve good results. Beall (2008) provided a general critique of full-text searching in the library catalogue, but this is the main criticism that lead to the bad reputation of ranking in this context.

The amount of text in each record in the catalogue varies greatly – from simple bibliographic data to a full-text document with hundreds of pages. Applying the same ranking algorithm to these types of records does not lead to good results. In contrast, records of each type should be ranked separately and then be put together into one results list.

*Popularity*

Popularity ranking can also be applied to library materials. Popularity could be measured on the basis of the individual item or on the basis of a group of items. For instance, a group can be built from all items by the same author, all items by the same publisher, or all items within a series.

We can apply factors using the number of items or the usage of an item (measures related to the circulation rate or the number of downloads). One can also take into account user ratings or citations when such data is available.

All this data is query-independent. Therefore, measures can be calculated periodically in advance, and only limited processing power is needed when calculating the results ranking for a certain query. Even if user ratings can be done continuously in the system, it will be sufficient to only update the popularity measures within a certain period of time.

*Freshness*

While freshness is the most-used ranking criterion in catalogues today, there is more to freshness than simply ordering results by date. It is hard to know when fresh items are particularly required, as the need for freshness may differ from one discipline to another. For example, fresh items may be crucial to a computer science researcher, but it may be a good idea to rely more heavily on text matching than on freshness for ranking items related to philosophy.

Therefore, it is important to determine the need for fresh items. This need for fresh items can be met either according to the circulation rate of an individual item – then it's more a measure of popularity – or by the circulation rates for items from a certain group. Such groups can be a broad discipline or even a specific subject heading. Again, the "need for freshness" factor is query-independent and can be calculated in advance.

*Locality*

Locality is a ranking factor that can take into account the physical location of the user as well as the availability of items in the results list. An item available at the local branch of the library could be ranked higher than items that are available only at a more distant branch. One can also use lending data to rank items. For some users, items not currently available for lending may be of little or no use and could therefore be ranked lower.

The physical location of the user can also be used in ranking. When a user is at home, we can assume that he will prefer to find electronic items that can be downloaded (Mercun & Zumer, 2008). When he is at the library, this restriction will not apply, and items available in print form can be ranked alongside electronic results. The location of the user can be determined through the IP address of his or her computer.

*Other ranking factors applicable to the library catalogue*

Adapting the groups of ranking factors used by the general-purpose Web search engines may not be enough, and there are many more ranking possibilities. The size or type of the item may be considered. Monographs may be favoured over edited books, books over journal articles (or vice versa), physical materials over online materials, etc.

User groups can also be taken into account. The needs of professors may differ greatly from the needs of undergraduate students, so different user groups may also determine ranking. Textbooks might be preferred in student searches, for instance.

Dividing library users into groups leads us to the question of personalization of results ranking. This requires individual usage data as well as click-stream data from navigation. However, collecting individual user data is always problematic and should be restricted to scenarios where the user knows what data is collected and has chosen this option. There are many ranking possibilities where anonymous statistical data (from the general user behaviour or from the behaviour of a certain group of users) can be used, so there is no real need for using individual user data.

A compilation of ranking factors suitable for library materials is one thing, but only a good combination of ranking factors can lead to good results. Decisions concerning a combination depend heavily on the individual collection and use cases.

**The composition of results lists**

The considerations reviewed so far should provide some idea of the complexity of ranking systems. However, our job does not end with applying these factors to the library's materials. Another problem arises in obtaining data from different sources. Apart from data coming directly from the catalogue, we also require circulation data (which was made anonymous) from the library system as well as location data, user data, and data from remote resources such as abstracts (or full texts) from the publishers.

While some sources are library-controlled, others are external and less easy to obtain. However, library-controlled collections also go far beyond the catalogue. Local digital repositories, course management systems, and the institution's Web sites may also be taken into account. Looking again at the origin of results in Web search engines, we can see that they face the same problems. Search results come from a variety of sources (databases such as Web, images, and news) and should be presented together in one results list (see Fig. 1). One can click on one of these results or click on the link that leads to the full results set of that particular database. This so-called "universal search" is a very good way to show users the diversity of results. While users often overlook the text links pointing to other databases, results from other databases that are injected into the main results lists are widely acknowledged.

How can this concept of universal search be applied to the library catalogue? One-box results, as shown in the screenshot, could point to databases licensed by the library or special collections built by the library.

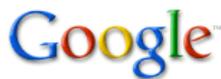
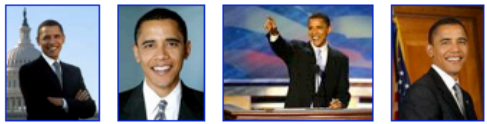
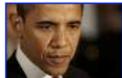

Fig. 1: Presentation of results from different databases within one results list (in Google)

When all of these things have been considered, one final challenge remains: the mixture of results on the results lists. Ranking algorithms in general prefer similar items. It could well be the case that results lists are crowded with "more of the same", but a certain variety would be useful. Search engines have become good at detecting duplicates and near-duplicates and removing them from their results lists. Additionally, specialised search engines such as Google Scholar group items (e.g., various instances of the same article). In some modern library search interfaces, this strongly needed step has already been taken.

Additionally, the breadth of the query should be taken into account. When a user types in a broad query, a good mixture of results will be expected. In the first few positions, a reference work, a textbook, a relevant journal, databases, and current items could be shown. However, for more specific queries, it would make sense to display an increased number of journal articles.

**Conclusion**

The main assumption of this article is that we are in a situation where the appropriate ranking factors for the library catalogue should be defined, while the implementation will not be the major problem. We must define what we want and not so much focus on the technical work. Some deep thinking is necessary on the "perfect results set" and how we can achieve it through ranking. We cannot expect computer scientists or information systems vendors to do this work for us. We know our material best and should therefore best know how a good results set could be achieved.

In conclusion, this paper argued that searching forms the core of the library catalogue. While other elements, such as usability or user guidance, must surely be considered, achieving relevant results is a key to the success of every library catalogue. The library catalogue cannot be "fixed" by adding modern, "2.0" features but only by improving the quality of the search results.

Secondly, good rankings can be achieved only through a combination of factors. Again, I want to stress that simple text matching is not enough. Thirdly, a good results list will always offer a mixture of results. There are some users (and some use-cases) that will require results listing all available literature on a given subject. However, in most cases, only a few relevant results are needed. The goal is to bring these results to the top of the list.

Looking into the future, I think that library catalogues will incorporate more of the attributes of a search engine than they do now. Ranking will be an integral part of this search engine. Most importantly, appreciation of the library catalogue as a search tool will depend on its ability to produce a good set of relevant results.

**Appendix**

| Group | Factor considered | Note |
|---|---|---|
| Text matching | Terms within bibliographic data, enriched data, full text | Bibliographic data does not contain enough text for a good term-based ranking. Great variety of the amount of text allows not applying the same ranking algorithm to all. |
|  | Field weighting |  |
|  | Availability of text | Considers whether additional text is available (e.g, reviews, TOC, full text). |
| Popularity | Number of local copies | Based on the individual item |
|  | Number of views | Based on the individual item |
|  | Circulation rate | Based on the individual item |
|  | Number of downloads | Based on the individual item |
|  | Author, publisher, book series, user ratings, citations | Based either on individual item or a group of items |
| Freshness | Publication date | Based on the individual item (could also be measured by its relationship to a group of items it belongs to) |
|  | Accession date | Based on the individual item (could also be measured by its relationship to a group of items it belongs to) |
| Locality | Physical location of the user (home, library, campus) | Location could be derived from IP address of a certain user. |
|  | Physical location of the item (central library, library branch, electronically available (i.e, no physical location important to the user) |  |
|  | Availability of item (available as a download, available at the library, currently not available) |  |
| Other | Size of item (e.g., number of pages) |  |
|  | Document type (monograph, edited book, journal article) | Could be related to the importance of certain document types within certain disciplines |
|  | User group (professors, undergraduate students, graduate students) |  |

---

[i] For instance, Google states that it uses more than 200 "signals", http://www.google.com/corporate/tech.html